\documentclass[12pt,a4paper]{article}
\usepackage{amssymb}
\usepackage{amsmath}
\usepackage[T2A]{fontenc}
\usepackage[cp1251]{inputenc}
\usepackage{epsfig,amscd,amssymb}

{ }

\begin{document}

\title{Singlet-triplet Hamiltonian for spin excitation in the
Kondo-insulator}

\author{A. F. Barabanov$^{a}$ and L.A. Maksimov$^{b}$\\
$^{a}$Institute for High Pressure Physics, Russian Academy
of Sciences, Troitsk, \\
 Moscow Region, 142190 Russia, e-mail: abarabanov@mtu-net.ru\\
 $^{b}$Kurchatov Institute Russian Research Center,
Moscow, 123182 Russia \\
}

\maketitle

\begin{abstract}
Within the framework of periodic asymmetric Anderson model for
Kondo isoulators an effective singlet-triplet Hamiltonian with
indirect antiferromagnetic $f-f$ exchange interaction is
introduced which allows to study analytically the dynamic magnetic
susceptibilities $\chi _{f}(\mathbf{k},\omega )$ of $\
f$-electrons. The approach allows to describe the three-level spin
excitation spectrum with a specific dispersion in $YbB_{12}$.
Distinctive feature of the consideration is the introduction of
small radius singlet and triplet collective $f-d$ excitations
which at movement on a lattice form low - and high-energy spin
bands.
\end{abstract}

\section{Introduction}
Kondo-insulators (KI) represent a special class of strongly
correlated systems. They demonstrate such phenomenon as
intermediate valency, a narrow gap (or a pseudo- gap) in a
spectrum of carriers of an order of $10$ $meV$\ which opens at low
temperatures, an unusual spectrum of spin fluctuations (with a
spin gap), a singlet ground state and a set of other interesting
properties (reviews \cite{Grewe1991}, \cite{Hewson1993},\
\cite{Aeppli1992}, \cite{Fisk1996}, \cite{Takabatake1998},
\cite{Degiorgi1999}, \cite{Riseborough2000}). The theory of KI is
based on the periodic Anderson model (PAM) which describes the
intrasite hybridization $\widehat{V}$ between $d-$ electrons and
localised $f$ -electrons with intrasite Coulomb $f-f$ repulsion
$U_{f}$. The intersite interaction answers to the hopping
Hamiltonian $\widehat{t}$\ for $d-$ electrons.
Considering
low-temperature spin fluctuations, we will discuss such a
characteristic KI as $YbB_{12}$. At $T<40K$, close to\textbf{\
}the spin gap\textbf{\ }edge the compound $YbB_{12}$\textbf{\
}demonstrates three dispersive excitations $M1$, $M2$, и $M3$ with
average energies $15$, $20$ и $40$ $meV$\ (see \cite{Alekseev07}
and references in \cite{Alekseev07}). The relaxation of low-energy
excitation $M1$ is suppressed, that allows to consider it as
resonance excitation. Peak $M1$ has a dispersion with a minimum,
and peak\ $M3$ with a maximum at $L$\ point which is the
aniferromagnetic (AFM) point for the face-centered cubic
$YbB_{12}$\ lattice.

Let's give the known accepted approaches to treatment of a magnetic
susceptibility $\chi (\mathbf{k},\omega )$ of $YbB_{12}$. The particle-hole
symmetric limit of the PAM with $f-$\ level filling $n_{f}=1$\ is often used
\cite{Grewe1991},\cite{Logan03} (however this mode is unjust in the mixed
valence regime (MVR) \cite{Logan07}). Sometimes strong $f-f$\ correlation is
considered in the slave boson mean-field approximation and $\chi
(\mathbf{k},\omega )$ calculations within the framework of PAM use random phase
approximation with introduction of crystalline electric field split for
$4f-$states \cite{Riseborough03}, \cite{Alekseev2004}, \cite{Fulde09}. And at
last, it was also proposed an alternative model which is explicitly based on
dielectric singlet ground state of $YbB_{12}$\ for which it is taken that
$Yb $\ -site $f^{14}$, $f^{13}d^{1}$\ configurations are nearly degenerate
\cite{Liu2-1}. Although all the treatments are devoted to the microscopic
origin and fine structure of the spin gap in $YbB_{12}$, in the majority of
cases $\chi (\mathbf{k},\omega )$\ considerations have a phenomenological
and qualitative character, are generally based on the single--impurity
Anderson model, so that AFM correlations between localized electrons arise
in higher order perturbation-theory approximations\ in hybridization and
hopping interactions. In particular, as far as we know, the explicit form
for $\chi (\mathbf{k},\omega )$ -dispersion of the lower spin excitation
branches$\ $was studied only in \cite{Fulde09}, \cite{Bar09}. However there
was no explanation of high-energy peak $M3$.

In present paper it will be investigated an asymmetrical PAM variant (APAM)
with two electrons per site in MVR $n_{f}\gtrsim 1$, assming that it is
realised in $YbB_{12}$.

In case of MVR the magnetic system actually does not have small parameter.
The close situation is realised in doped $CuO_{2}$\ planes in HTSC cuprates.
where the indirect \ $d-d$\ intersite AFM exchange is of the same order as
$p $-holes interaction with the spin subsystem. In this case it is known,
that for the adequate account of strong correlations for the low-energy
excitations first place is necessary to solve a problem for small radius
cluster, next to construct low-energy Zhang-Rice polaron (collective
excitation of small radius) \cite{ZhangRice1988} (see also
\cite{MaksHaBar1998}) and only in the end to consider such a polaron movement on
a lattice.

Similarly, for the accepted model it will be shown, that observable
three-level spin excitation character can be described if from the very
beginning the problem is considered on the basis of one- and two- site $f-d$
collective triplet-singlet excitations (small radius excitations).
Consideration differs from the previous themes, that $\chi
(\mathbf{k},\omega )$\ is defined exactly\ by movement of these excitations against a
singlet ground state (movement is determined by\ the indirect AFM $f$--$f$
exchange).

Below we will be interested only in a susceptibility$\chi
_{f}(\mathbf{k},\omega )$ of $f$-- electrons assuming that at low temperatures it is main
for full $\chi (\mathbf{k},\omega )$. The effective singlet-triplet
Hamiltonian $\widehat{H}_{s-t}$ is itroduced for two site-cluster analysis
and $\chi _{f}(\mathbf{k},\omega )$ calculations. It is substantiated that
$\widehat{H}_{s-t}$ describes the basic properties of APAM.

Initial background of our approach are close to the phenomenological $\chi
(\mathbf{k},\omega )$ consideration in \cite{Liu2-1}, however as a whole the
approach is more realistic and it allows not only to explain $M1,$ $M2,$ $M3$
peaks, but also to find their spectrum.

\section{Effective Hamiltonian and two-site cluster.}

As shown in \cite{Bar09} for the description of KI spin subsystem with
$n_{f}\gtrsim 1$ it is convenient to use the effective Hamiltonian
$\widehat{H}_{J}$ which turns out from PAM Hamiltonian by a replacement of a hopping
turm $\widehat{t}$\ on AFM $f-f$ indirect exchange interaction
$\widehat{J}$\ on the nearest neighbors.

In conventional\ notation $\widehat{H}_{J}$ Hamiltonian has the form

\begin{eqnarray}
\widehat{H}_{J} &=&\widehat{H}_{0}+\widehat{J};\;\ \ \
\widehat{J}=\frac{1}{2}J\sum_{\mathbf{r,g}}\widehat{\mathbf{S}}_{f,\mathbf{r}}\widehat{\mathbf{S}}_{f,\mathbf{r+g}},\quad \quad
\widehat{H}_{0}=\sum_{\mathbf{n}}\widehat{H}_{0,\mathbf{n}};\   \label{h1} \\
\widehat{H}_{0,\mathbf{n}} &=&[\sum_{\sigma }\varepsilon
_{f}\widehat{f}_{\mathbf{n,}\sigma }^{+}\widehat{f}_{\mathbf{n,}\sigma
}+U_{f}\widehat{f}_{\mathbf{n,}+}^{+}\widehat{f}_{\mathbf{n,}+}\widehat{f}_{\mathbf{n,}-}^{+}\widehat{f}_{\mathbf{n,}-}]+\sum_{\sigma }\varepsilon
_{d}\widehat{d}_{\mathbf{n,}\sigma }^{+}\widehat{d}_{\mathbf{n,}\sigma
}+\widehat{V}_{\mathbf{n}};  \notag \\
\quad \widehat{V}_{\mathbf{n}} &=&V\sum_{\sigma
}(\widehat{f}_{\mathbf{n,}\sigma }^{+}\widehat{d}_{\mathbf{n,}\sigma }+\widehat{d}_{\mathbf{n,}\sigma
}^{+}\widehat{f}_{\mathbf{n,}\sigma });\quad \text{ \ \ }\sigma =\pm ,
\notag
\end{eqnarray}

$\mathbf{g}$ is a nearest neighbor vector. Here we neglect orbital
degeneracy. $V$ and $t\ $- amplitudes of $\widehat{V}$\ и $\widehat{t}$
interactions; $\varepsilon _{f}$ and $\varepsilon _{d}$ are energies of
$f-$and $d-$levels, below we put $\varepsilon _{d}=0;\ J$ - a constant of AFM
$f-f$ exchange. MVR with $n_{f}\gtrsim 1$ is characterized by the following
energy parametres relations: $U_{f}+\varepsilon _{f}\simeq V>0$\ and $V\ll
U_{f}\simeq -\varepsilon _{f}$.

Let us discuss a full set of two-electron eigen states of one-site
Hamiltonian$\ \widehat{H}_{0,\mathbf{r}}$\ (\ref{h1})$\ $in order to
validate an introduced below effective singlet-triplet Hamiltonian
$\widehat{H}_{s-t}$ (it will differ\ from (\ref{h1}) by\ of an intrasite term).

At $T<-\varepsilon _{f}$\ and the accepted parametres it is possible at once
to\ omit а high -energy state state
$2_{d\mathbf{r}}=\widehat{d}_{\mathbf{r,}+}^{+}\widehat{d}_{\mathbf{r,}-}^{+}|0\rangle $\ ($|0\rangle $-
site vacuum) with two particles on $d$ - level.

There are one singlet $\phi _{\mathbf{r}}$\ and three triplet states with
one electron on $f-$\ and $d-$levels.\ Without the hybridization account
$\phi _{\mathbf{r}}$\ and $\psi _{i=0,\pm ;\mathbf{r}}$\ are degenerate (with
energy $\varepsilon _{f}<0$) and are expessed as

\begin{eqnarray*}
\phi _{\mathbf{r}} &=&\widehat{\phi }_{\mathbf{r}}^{+}|0\rangle ,\ \ \
\widehat{\phi
}_{\mathbf{r}}^{+}=\frac{1}{\sqrt{2}}(\widehat{f}_{\mathbf{r}+}^{+}\widehat{d}_{\mathbf{r}-}^{+}-\widehat{f}_{\mathbf{r}-}^{+}\widehat{d}_{\mathbf{r}+}^{+});\ \psi _{\pm 1,\mathbf{r}}=\widehat{\psi }_{\pm
1,\mathbf{r}}^{+}|0\rangle ,\ \ \widehat{\psi }_{\pm
1,\mathbf{r}}^{+}=\widehat{f}_{\mathbf{r,}\pm }^{+}\widehat{d}_{\mathbf{r,}\pm }^{+}.; \\
\psi _{0,\mathbf{r}} &=&\widehat{\psi }_{0,\mathbf{r}}^{+}|0\rangle ,\
\widehat{\psi
}_{0,\mathbf{r}}^{+}=\frac{1}{\sqrt{2}}(\widehat{f}_{\mathbf{r}+}^{+}\widehat{d}_{\mathbf{r}-}^{+}+\widehat{f}_{\mathbf{r}-}^{+}\widehat{d}_{\mathbf{r}+}^{+});\
\end{eqnarray*}

Hybridization with a singlet two $f$\ -electrons state
$2_{f\mathbf{r}}=\widehat{2}_{f\mathbf{r}}^{+}|0\rangle
,~~\widehat{2}_{f\mathbf{r}}^{+}=\widehat{f}_{\mathbf{r,}+}^{+}\widehat{f}_{\mathbf{r,}-}^{+}|0\rangle $\
(its energy $2\varepsilon _{f}+\ U_{f}$) removes degeneration between a
singlet $\phi _{\mathbf{r}}$\ and triplets $\psi _{i=0,\pm ;\mathbf{r.}}$
Triplet states don't hybridize with singlets. As a result of hybridization
between $\phi _{\mathbf{r}}$\ and $2_{f\mathbf{r}}$\ two singlet eigen
states of $\widehat{H}_{0,\mathbf{r}}$\ have the form

\begin{equation*}
\varphi _{\mathbf{r;\pm }}=\widehat{\varphi }_{\mathbf{r;\pm }}^{+}|0\rangle
,~~\ \widehat{\varphi }_{\mathbf{r;\pm }}^{+}=A_{\mathbf{\pm }}\widehat{\phi
}_{\mathbf{r}}^{+}+B_{\mathbf{\pm
}}\widehat{2}_{f\mathbf{r}}^{+};~~E_{\varphi \mathbf{;\pm }}=\frac{1}{2}\{3\varepsilon _{f}+U_{f}\pm
\sqrt{(\varepsilon _{f}+U_{f})^{2}+8V^{2}}\}
\end{equation*}

At $U_{f}+\varepsilon _{fd}\simeq V$\ we have $E_{\varphi }=E_{\varphi
;-}\simeq \varepsilon _{f}-V,$\ $E_{\varphi ;+}\simeq \varepsilon
_{f}+2V$,$\ E_{\psi }=\varepsilon _{f}$. The lower singlet state $\varphi
_{\mathbf{r}}\equiv \varphi _{\mathbf{r;-}}=A\widehat{\phi }_{\mathbf{r}}^{+}|0\rangle
+B\widehat{2}_{f\mathbf{r}}^{+}|0\rangle $ is a ground one. Singlet-triplet
transitions will be considered at $T<V$, then the upper singlet state
$\varphi _{\mathbf{r;+}}$\ can be omitted ($E_{\varphi ;+}-E_{\varphi }\simeq
3V$). Thus the hybridization role in one-site Hamiltonian$\
\widehat{H}_{0,\mathbf{r}}$\ is reduced to the energy renormalization\ of a ground singlet
so that $E_{\varphi }<E_{\psi }=\varepsilon _{f}$.

Below it appears, that instead of triplet states $\psi _{i=0,\pm
;\mathbf{r}} $\ it is more convenient to use basis \cite{Landau1989}

\begin{equation*}
\begin{array}[t]{c}
w_{z}=\psi _{0};~~~~w_{x}=\frac{1}{\sqrt{2}}(\psi _{-1}-\psi _{1});~~~\
w_{y}=\frac{i}{\sqrt{2}}(\psi _{-1}+\psi _{1}),\end{array}\end{equation*}

which explicitly reflects a spherical symmetry of a problem.

In considered approximation the basis of states $\varphi _{\mathbf{r}}$,
$w_{q,\mathbf{r}}$ ($q=x,y,z$) is a full one-site basis. In this bases
$\widehat{H}_{0}$\ (\ref{h1})$\ $takes a diagonal form $\widehat{H}_{0,s-t}$
and $\widehat{H}_{J}$ is transformed in singlet-triplet Hamiltonian$\
\widehat{H}_{s-t}$\

\begin{equation}
\widehat{H}_{s-t}=\widehat{H}_{0,s-t}+\widehat{J};~~\ \ \ \
\widehat{H}_{0,s-t}=e\sum_{\mathbf{r,}q}\widehat{Z}_{\mathbf{r}}^{qq};\text{ }
\label{h2}
\end{equation}\

here and below $\widehat{Z}_{\mathbf{r}}^{\lambda \lambda ^{\prime }}$\ are
Hubbard projection operators into corresponding states, $\lambda $, $\lambda
^{\prime }=$\ $\varphi $, $w_{q}$ ($\widehat{Z}-$operators indexes
$q=x,y,z$\ correspond to $w_{q}$); $e=E_{\psi }-E_{\varphi },$ below energy will
counted from the ground state energy $E_{\varphi }$. At characteristic
relation between parameters $\varepsilon _{f}+\ U_{f}=2V$\ the triplet
energy is $e=0.7V$, and $\varphi _{\mathbf{r}}$\ -structure $\varphi
_{\mathbf{r}}=A\phi _{\mathbf{r}}+B2_{f\mathbf{r}}$\ answers the values
$A^{2}\simeq 0.8,$ $\ B^{2}\simeq 0.2$. At $V\simeq 30$ $meV$ we have for $e$
-value $e\simeq 20$ $meV,$\ this value is close to \ $M1$, $M2$ -peaks
energy.

In the basis $\ \varphi _{\mathbf{r}}$, $w_{q,\mathbf{r}}$\ $f-$spin site
operator $\widehat{S}_{f\mathbf{r}}^{q}$ ($q=x,y,z$) looks like

\begin{equation}
\begin{array}{c}
\widehat{S}_{f\mathbf{r}}^{q}=\frac{1}{2}(A\widehat{Z}_{\mathbf{r}}^{\varphi
q}+A\widehat{Z}_{\mathbf{r}}^{q\varphi }+i\varepsilon _{qq^{\prime
}q^{\prime \prime }}\widehat{Z}_{\mathbf{r}}^{q^{\prime \prime }q^{\prime
}}).\end{array}
\label{s1}
\end{equation}

In what follows in the expression (\ref{s1})\textbf{\ }we will not
distinguish the states $\varphi _{\mathbf{r}}$ и $\phi _{\mathbf{r}}$
and will assume for the simplicity $A=1.$

For a two-site cluster with the Hamiltonian\ $\widehat{H}_{0,s-t}$\
(\ref{h2}) let us construct eigen states which have eigen values $W$\ for a cluster
full spin operater. The analysis of these states allows to clarify the
necessity of falling outside one-site approximation. The basis of four
one-site operaters $\widehat{\phi }_{\mathbf{r}}^{+}$,
$\widehat{w}_{q,\mathbf{r}}^{+}$ gives $16$ eigen states (corresponding energies are
designated as $E^{0})$ with\ $W=0,1,2$.

There are two singlet states $\Phi _{\mathbf{n,m}}$\ and $X_{\mathbf{n,m}}$
which have a form $\Phi _{\mathbf{n,m}}\mathbf{=}\widehat{\phi
}_{\mathbf{n}}^{+}\widehat{\phi }_{\mathbf{m}}^{+}|0\rangle $,
$X_{\mathbf{n,m}}$\textbf{=}$\frac{1}{\sqrt{3}}\sum_{q}\widehat{w}_{q,\mathbf{n}}^{+}\widehat{w}_{q,\mathbf{m}}^{+}|0\rangle $ (here $|0\rangle $-cluster vucuum) with energies
$E_{\Phi }^{0}=0$,\ $E_{X}^{0}=2e$.

Six triplet states are
$B_{1}^{q}=\frac{1}{\sqrt{2}}(1-\widehat{T}_{\mathbf{n,m}})\widehat{w}_{q,\mathbf{n}}^{+}\widehat{\phi
}_{\mathbf{m}}^{+}|0\rangle ;$ $\
D_{1}^{q}=\frac{1}{\sqrt{2}}(1+\widehat{T}_{\mathbf{n,m}})\widehat{w}_{q,\mathbf{n}}^{+}\widehat{\phi }_{\mathbf{m}}^{+}|0\rangle $,
here and below $\widehat{T}_{\mathbf{n,m}}$\ is an operator of $\mathbf{n},$
$\mathbf{m}$\ -sites permutation. The states $B_{1}^{q}$ и $D_{1}^{q}$
have the same energy $E_{B_{1}^{q}(D_{1}^{q})}^{0}=e$ but different parity
$T $\ relative to $\widehat{T}_{\mathbf{n,m}}$ ($T_{B_{1}^{q}}=-1$,
$T_{D_{1}^{q}}=+1$).

Three more triplet states have the form $\begin{array}{c}
B_{2}^{q}=\ \frac{1}{\sqrt{2}}i\varepsilon _{qq^{\prime }q^{\prime \prime
}}w_{q^{\prime },\mathbf{n}}^{+}w_{q^{\prime \prime
},\mathbf{m}}^{+}|0\rangle\end{array}$\ with $E_{B_{2}^{q}}^{0}=2e$\ and $T_{B_{2}^{q}}=-1$.

At last there are five quintet\ states $L^{l}$\ , $l=1\div 5$\ with full
spin $W=2,$\ $E_{L^{l}}^{0}=2e$, parity $T=(+1)$, they are described by the
following wave functions

$\begin{array}{c}
\frac{1}{\sqrt{2}}(1+\widehat{T}_{\mathbf{n,m}})w_{x,\mathbf{n}}^{+}w_{y,\mathbf{m}}^{+};\ \text{\
}\frac{1}{\sqrt{2}}(1+\widehat{T}_{\mathbf{n,m}})w_{y,\mathbf{n}}^{+}w_{z,\mathbf{m}}^{+};\text{ }\
\frac{1}{\sqrt{2}}(1+\widehat{T}_{\mathbf{n,m}})w_{z,\mathbf{n}}^{+}w_{x,\mathbf{m}}^{+};\text{ \
} \\
\frac{1}{\sqrt{2}}(w_{x,\mathbf{n}}^{+}w_{x,\mathbf{m}}^{+}-w_{y,\mathbf{n}}^{+}w_{y,\mathbf{m}}^{+})|0\rangle ;\ \text{\
}\frac{1}{\sqrt{6}}(w_{x,\mathbf{n}}^{+}w_{x,\mathbf{m}}^{+}+w_{y,\mathbf{n}}^{+}w_{y,\mathbf{m}}^{+}-2w_{z,\mathbf{n}}^{+}w_{z,\mathbf{m}}^{+})|0\rangle
.\end{array}$

The states with $T=(+1)$ correspond to Fourier transform with
$\mathbf{k}=(0),$ $T=(-1)$ with $\mathbf{k}=(\pi )$, the last one is an anolog of AFM
point.

The exchange $\widehat{J}$\ (\ref{h1})\ leads to mixing between the states
with coinciding $T$\ and $W$ (and coinsiding $q$\ for triplets). The matrix
of the Hamiltonian in $\Phi _{\mathbf{n,m}}$\ ,\ $X_{\mathbf{n,m}}$\ basis
gives a cluster ground state $\Psi _{\mathbf{n,m}}$. It is formed mainly by
$\Phi _{\mathbf{n,m}}$\ which is factorable on $\mathbf{n,m}$. $\Psi
_{\mathbf{n,m}}$ contains an admixture of $X_{\mathbf{n,m}}$\ state (with an
amplitude proportional to $J/e$ at small $J$), $X_{\mathbf{n,m}}$\ state is
not factorable. An explicit $\Psi _{\mathbf{n,m}}$-form gives a possibility
of simpliest calculations for two-site correlation functions at low $T$.

The main result of the above classification is a construction of two-site
nonfactorable on sites $B_{2}^{q}$\ triplet states. They are the only states
that permit singlet $\Psi _{\mathbf{n,m}}$-- triplet excitation with the
energy $E_{B_{2}^{q}}^{0}=2e$. Below it is shown that exactly the movement
of this two-site excitation leads to a spin branch which answers the
position and dispersion of $M3$ peak.

\section{Spin susceptibility of $f-$electrons.}

\ For the description of a real spin spectrum it is necessary to turn from a
two-site cluster problem to a consideration of an infinite lattice with the
Hamiltonian $\widehat{H}_{s-t}$.(\ref{h2}). The susceptibility $\chi
_{f}(\mathbf{k},\omega )$ is given by the two-time spin Green's function (GF)

$G_{f}^{z}(\mathbf{k},\omega )=\langle
\widehat{S}_{f,\mathbf{k}}^{z}|\widehat{S}_{f,\mathbf{-k}}^{z}\rangle _{\omega }=-\chi
_{f}(\mathbf{k},\omega )$, \
$\widehat{S}_{f,\mathbf{k}}^{z}=\frac{1}{\sqrt{N}}{\,}\sum\limits_{\mathbf{r}}e^{-i\mathbf{kr}}\widehat{S}_{f,\mathbf{r}}^{z}.$

Since $YbB_{12}$ lacks long range spin order we describe the properties of
its spin system by a spherically symmetric approach (SSA). similar to that
used in \cite{Shimahara91}. Within the limits SSA
$G_{f}^{x(y)}(\mathbf{k},\omega )=G_{f}^{z}(\mathbf{k},\omega )$ and the following averages are
equal to zero: $\langle \widehat{Z}_{\mathbf{r}}^{\phi q}\rangle =0;$\
$\langle \widehat{Z}_{\mathbf{r}}^{q^{\prime }q^{\prime \prime }}\rangle =0$\
at $q^{\prime }\neq q^{\prime \prime }$.

Taking into account (\ref{s1})\ the site representation of
$G_{f}^{z}(\mathbf{k},\omega )$ breaks up into a sum of terms $\langle
\widehat{Z}_{\mathbf{n}}^{\phi z}|\widehat{Z}_{\mathbf{r}}^{z\phi }\rangle _{\omega }$,
$\langle \widehat{Z}_{\mathbf{n}}^{z\phi }|\widehat{Z}_{\mathbf{r}}^{\phi
z}\rangle _{\omega })$, $\langle \widehat{Z}_{\mathbf{n}}^{\phi
z}|\widehat{Z}_{\mathbf{r}}^{\phi z}\rangle _{\omega }$,$\ \langle
\widehat{Z}_{\mathbf{n}}^{z\phi }|\widehat{Z}_{\mathbf{r}}^{z\phi }\rangle _{\omega }$, $\langle
i\varepsilon _{zq^{\prime }q^{\prime \prime
}}\widehat{Z}_{\mathbf{n}}^{q^{\prime \prime }q^{\prime }}|i\varepsilon
_{zq_{1}q_{2}}\widehat{Z}_{\mathbf{r}}^{q_{1}q_{2}}\rangle _{\omega }$, $\langle
\widehat{Z}_{\mathbf{n}}^{\phi z}|i\varepsilon
_{zq_{1}q_{2}}\widehat{Z}_{\mathbf{r}}^{q_{1}q_{2}}\widehat{T}_{\mathbf{r}}^{z}\rangle _{\omega }$, $\langle i\varepsilon
_{zq^{\prime }q^{\prime \prime }}\widehat{Z}_{\mathbf{n}}^{q^{\prime \prime
}q^{\prime }}|\widehat{Z}_{\mathbf{r}}^{z\phi }\rangle _{\omega }$, $\langle
\widehat{Z}_{\mathbf{n}}^{z\phi }|i\varepsilon
_{zq_{1}q_{2}}\widehat{Z}_{\mathbf{r}}^{q_{1}q_{2}}\rangle _{\omega }$, $\langle i\varepsilon
_{zq^{\prime }q^{\prime \prime }}\widehat{Z}_{\mathbf{n}}^{q^{\prime \prime
}q^{\prime }}|\widehat{Z}_{\mathbf{r}}^{z\phi }\rangle _{\omega }$.

As mentioned above the ground state is formed mainly by one-site singlets
$\phi _{\mathbf{r}}$. Then spin excitation first of all should be described
by one-site singlet-triplet operator $\widehat{Z}_{\mathbf{r}}^{z\phi }$\
with energy $e$. Therefore below we will begin the discussion from the GF
$\widetilde{G}_{\mathbf{n-r}}^{z}(\omega )=\langle
\widehat{Z}_{\mathbf{n}}^{\phi z}|\widehat{Z}_{\mathbf{r}}^{z\phi }\rangle _{\omega }$ which
corresponds to the first term of full GF $G_{f}^{z}(\mathbf{k},\omega )$\
and has obviously the basic pole close to $e$.

The equation for GF $\langle \widehat{Z}_{\mathbf{n}}^{\phi
z}|\widehat{Z}_{\mathbf{r}}^{z\phi }\rangle _{\omega }$ has the form:

\begin{eqnarray}
(z-e)\langle \widehat{Z}_{\mathbf{n}}^{\phi
z}|\widehat{Z}_{\mathbf{r}}^{z\phi }\rangle _{\omega } &=&\delta _{\mathbf{n,r}}P_{\phi w}+\langle
\lbrack \widehat{Z}_{\mathbf{n}}^{\phi
z};\widehat{J}]|\widehat{Z}_{\mathbf{r}}^{z\phi }\rangle _{\omega }; \\
\text{\textbf{\ }}P_{\phi w} &=&P_{\phi }-P_{w};\ \text{\ \ }P_{\phi
}=\langle \widehat{Z}_{\mathbf{r}}^{\phi \phi }\rangle ;\text{ \
}P_{w}=\langle \widehat{Z}_{\mathbf{r}}^{qq}\rangle \ .
\end{eqnarray}

Carrying out the commutation $[\widehat{Z}_{\mathbf{n}}^{\phi
z};\widehat{J}] $\ in an explicit form with the account of substitution of expressions
(\ref{s1}) in exchange interaction $\widehat{J}$, we will receive the
following equation for $\langle \widehat{Z}_{\mathbf{n}}^{\phi z}|\rangle
\equiv \langle \widehat{Z}_{\mathbf{n}}^{\phi
z}|\widehat{Z}_{\mathbf{r}}^{z\phi }\rangle _{\omega }$\ (here and in obvious cases a simplified GF
notation is used)

\begin{equation}
\begin{array}[t]{c}
(z-e)\langle \widehat{Z}_{\mathbf{n}}^{\phi z}|\rangle =\delta
_{\mathbf{n,r}}P_{\phi w}+\frac{J}{8}\sum_{\mathbf{g}}\{\langle
\widehat{R}_{\mathbf{n;g}}|\rangle \text{+}[(\langle \widehat{V}_{\mathbf{n;g}}|\rangle -\langle
\widehat{W}_{\mathbf{n;g}}|\rangle )\text{+} \\
\text{+}2(\langle \widehat{U}_{\mathbf{n;g}}|\rangle \text{-}\langle
\widehat{V}_{\mathbf{ng;-g}}|\rangle )]\text{+}\langle
\widehat{D}_{\mathbf{n;g}}|\rangle \text{+}\langle \widehat{L}_{\mathbf{n;g}}|\rangle -\langle
\widehat{N}_{\mathbf{n;g}}|\rangle \text{+}2[\langle
\widehat{U}_{\mathbf{n;g}}^{+}|\rangle -\langle \widehat{W}_{\mathbf{ng;-g}}^{+}|\rangle \
]\}.\text{\ }\end{array}
\label{eq2}
\end{equation}

Here the arisen two-site operators have the form

\begin{equation}
\begin{array}[t]{c}
\ \widehat{U}_{\mathbf{n;g}}\text{=}\widehat{Z}_{\mathbf{n+g}}^{\phi
z}\widehat{Z}_{\mathbf{n}}^{\phi \phi };\
\widehat{V}_{\mathbf{n;g}}=\widehat{Z}_{\mathbf{n+g}}^{dz}\,\widehat{Z}_{\mathbf{n}}^{\phi d};\ \
\widehat{W}_{\mathbf{n;g}}=\widehat{Z}_{\mathbf{n+g}}^{zd}\widehat{Z}_{\mathbf{n}}^{\phi
d};\ \widehat{R}_{\mathbf{n;g}}=i\varepsilon
_{zuq}\widehat{Z}_{\mathbf{n+g}}^{\phi u}\widehat{Z}_{\mathbf{n}}^{\phi q};\text{ \ \ \ \ } \\
\ \widehat{D}_{\mathbf{n;g}}=i\,\varepsilon
_{zuq}\widehat{Z}_{\mathbf{n+g}}^{u\phi }\widehat{Z}_{\mathbf{n}}^{\phi
q};\;\widehat{L}_{\mathbf{n;g}}=i\varepsilon
_{zuq}\widehat{Z}_{\mathbf{n+g}}^{qu}\widehat{Z}_{\mathbf{n}}^{\phi \phi };\;\widehat{N}_{\mathbf{n;g}}=i\varepsilon
_{duq}\widehat{Z}_{\mathbf{n+g}}^{qu}\widehat{Z}_{\mathbf{n}}^{dz}.\end{array}
\label{not5}
\end{equation}

The GF in the right part of (\ref{eq2}) describe transitions with energy
$\omega \approx -e$ (operators $\widehat{U}_{\mathbf{n;g}}^{+}$,
$\widehat{W}_{\mathbf{ng;}\overline{\mathbf{g}}}^{+}$), $\omega \approx 0$\ (operators
$\widehat{D}_{\mathbf{n;g}}$, $\widehat{L}_{\mathbf{n;g}}$,
$\widehat{N}_{\mathbf{n;g}}$)$,$ $\omega \approx e\ $(operators
$\widehat{U}_{\mathbf{n;g}} $, $\widehat{V}_{\mathbf{n;g}}$, $\widehat{W}_{\mathbf{n;g}}$\ )$\ $и
$\omega \approx 2e$ (operator $\widehat{R}_{\mathbf{n;g}}$). Let us clarify
their physical meaning.

The GF with $\widehat{D}_{\mathbf{n;g}}$\ , $\widehat{L}_{\mathbf{n;g}}$\ ,
$\widehat{N}_{\mathbf{n;g}}$\ operators answer triplet-triplet against a
singlet background, they should lead to a quasielastic peak which we do not
consider. In the simpiest mean field approach it is possible to neglect also
the GF $\langle \widehat{U}_{\mathbf{n;g}}^{+}|\rangle ~,~\langle
\widehat{W}_{\mathbf{ng;}\overline{\mathbf{g}}}^{+}|\rangle $ as they.do not give a\
return to the initiale GF $\langle \widehat{Z}_{\mathbf{i}}^{\phi z}|\rangle
$.

The operators $\widehat{U}_{\mathbf{n;g}}$, $\widehat{V}_{\mathbf{n;g}}$,
$\widehat{W}_{\mathbf{n;g}}$\ (\ref{eq2}) cjrrespond to the same energy
transitions as $\widehat{Z}_{\mathbf{n}}^{\phi z}$, $\omega =e$. It should
be noted that the term $\langle \widehat{U}_{\mathbf{n;g}}|\rangle $\
describes a triplet excitation movement from site $\mathbf{n}$\ to a
neighbouring $\mathbf{n+g}$. Operators $\widehat{U}_{\mathbf{n;g}}$,
$\widehat{V}_{\mathbf{n;g}}$, $\widehat{W}_{\mathbf{n;g}}$ permit a mean
field approach with a\ return to GF $\langle \widehat{Z}_{\mathbf{i}}^{\phi
z}|\rangle $: $[(\langle \widehat{V}_{\mathbf{n;g}}|\rangle -\langle
\widehat{W}_{\mathbf{n;g}}|\rangle )$ +$2(\langle
\widehat{U}_{\mathbf{n;g}}|\rangle $-$\langle \widehat{V}_{\mathbf{ng;-g}}|\rangle )]$ $\approx
P_{\phi w}\langle \widehat{Z}_{\mathbf{n+g}}^{\Phi
z}|\widehat{Z}_{\mathbf{r}}^{z\Phi }\rangle $. In this approximation (\ref{eq2})\ takes a form

\begin{equation}
\begin{array}{c}
\ (z-e)\langle \widehat{Z}_{\mathbf{n}}^{\phi z}|\rangle =\delta
_{\mathbf{n,r}}P_{\phi w}+\frac{J}{8}\sum_{\mathbf{g}}\{\langle
\widehat{R}_{\mathbf{n;g}}|\rangle +2P_{\phi w}\langle \widehat{Z}_{\mathbf{n+g}}^{\phi
z}|\rangle \}\text{ .}\end{array}
\label{eq3}
\end{equation}

In a momentum\ representation (\ref{eq3}) gives

\begin{equation}
\lbrack z-e-\frac{J}{4}P_{\phi
w}z_{l}g_{\mathbf{k}}]\widetilde{G}_{\mathbf{k}}^{z}(\omega )=P_{\phi w}+\frac{J}{8}\sum_{\mathbf{g}}\langle
\widehat{R}_{\mathbf{k;g}}|\rangle ,\text{ \ }\widetilde{G}_{\mathbf{k}}^{z}(\omega
)=\langle \widehat{Z}_{\mathbf{k}}^{\phi z}|\widehat{Z}_{-\mathbf{k}}^{z\phi
}\rangle \text{\ ,}  \label{eq3_k}
\end{equation}

here $z_{l}$\ -number of the nearest neighbours,\
$g_{\mathbf{k}}=z_{l}^{-1}\sum_{\mathbf{g}}e^{i\mathbf{kg}}$. For the face-centered cubic
$YbB_{12}$\ lattice \ $z_{l}=12;$ $g_{\mathbf{k}}=\frac{1}{3}[\cos k_{x}\cos
k_{y}+\cos k_{y}\cos k_{z}+$ $\cos k_{z}\cos k_{x}]$, the AFM vector is
$\mathbf{Q=}\frac{\pi }{2}[1;1;1]$.\

If to neglect the GF $\langle \widehat{R}_{\mathbf{k;g}}|\rangle $ in the
equation (\ref{eq3_k}) it will describe a triplet excitation spectrum spin
$E_{\mathbf{k;}1}^{0}=e+\frac{J}{4}P_{\phi w}z_{l}g_{\mathbf{k}}$\ with a gap
equal $e$ and a dispersion part $\frac{J}{4}P_{\phi w}z_{l}g_{\mathbf{k}}$
with minimum at $Q$, as well as is observed for branch $M1$.

If in (\ref{eq3_k}) we neglect the GF $\langle
\widehat{R}_{\mathbf{k;g}}|\rangle $ but do not restrict ourself by a mean field approximation for
$\widehat{U}_{\mathbf{n;g}}$, $\widehat{V}_{\mathbf{n;g}}$,
$\widehat{W}_{\mathbf{n;g}}$\ -operators then the equations of motion must be introduced
for the corresponding GF $\langle \widehat{U}_{\mathbf{n;g}}|\rangle $,
$\langle \widehat{V}_{\mathbf{n;g}}|\rangle $, $\langle
\widehat{W}_{\mathbf{n;g}}|\rangle $. We do not represent these equations because of their
cumbersome form. Nevertheless it may be seen that their account leads to
additional branches of a spectrum close to a pole $\omega \simeq e$.
Effectively these branches correspond to $M1$ peak broadening with some $J$
- scale structure. The last can be treated as analogue of peak $M2$.

The account of a term $\langle \widehat{R}_{\mathbf{n;g}}|\rangle $\ in the
equations (\ref{eq2}), (\ref{eq3_k})\ is important on principle. The
operator $\ \widehat{R}_{\mathbf{n;g}}=i\varepsilon
_{zuq}\widehat{Z}_{\mathbf{n+g}}^{\phi u}\widehat{Z}_{\mathbf{n}}^{\phi q}$ answers to
singlet-triplet transition between two-site singlet state $\Phi
_{\mathbf{n,m}}$\ and a two-site triplet state $B_{2}^{z}$ with energy $\omega =2e$.\
This transition lead to $\langle \widehat{Z}_{\mathbf{n}}^{\phi
z}|\widehat{Z}_{\mathbf{r}}^{z\phi }\rangle $\ GF pole close to $M3-$peak energy
$40meV~~$if $e\simeq 20~meV$.

The equation for $\langle \widehat{R}_{\mathbf{n;g}}|\rangle $ has a form

\begin{equation}
\begin{array}[t]{c}
(z-2e)\sum_{\mathbf{g}}\langle \widehat{R}_{\mathbf{n;g}}|\rangle
\text{=}K_{\widehat{R}_{\mathbf{n}};\widehat{Z}_{\mathbf{r}}^{z\phi
}}+\sum_{\mathbf{g}}\text{\textbf{\{}}\frac{J}{4}\{2(\langle \widehat{U}_{\mathbf{ng}}|\rangle
-\langle \widehat{U}_{\mathbf{n}}|\rangle \mathbf{)}-\langle
\widehat{R}_{\mathbf{n;g}}|\rangle -\langle \widehat{M}_{\mathbf{n;g}}|\rangle
\mathbf{\}}\text{\textbf{+}} \\
\text{\textbf{+}}4J\{\mathbf{\sum_{\mathbf{b}}\overline{\Delta
}_{\mathbf{b;g}}}[-C_{ZZ,g}^{x}\langle \widehat{Z}_{\mathbf{n+b}}^{\phi z}|\rangle
\text{+}C_{ZZ,|\mathbf{g-b}|}^{x}\langle \widehat{Z}_{\mathbf{n}}^{\phi z}|\rangle
]\mathbf{+}\sum_{\mathbf{b}}\overline{\Delta
}_{\mathbf{b;}\overline{\mathbf{g}}}[C_{ZZ,g}^{x}\langle \widehat{Z}_{\mathbf{n+g+b}}^{\phi z}|\rangle
\text{\textbf{-}}C_{ZZ,|\mathbf{g+b}|}^{x}\langle \widehat{Z}_{\mathbf{n+g}}^{\phi
z}|\rangle ]\text{\}}\mathbf{;} \\
K_{\widehat{R}_{\mathbf{n}};\widehat{Z}_{\mathbf{r}}^{z\phi
}}\text{=}\sum_{\mathbf{g}}\langle \lbrack
\widehat{R}_{\mathbf{n;g}};\widehat{Z}_{\mathbf{r}}^{z\phi }]\rangle ~;C_{ZZ,l}^{x}=\langle \widehat{Z}_{\mathbf{n+l}}^{\phi
x}\widehat{Z}_{\mathbf{n}}^{x\phi }\rangle ;~~~\overline{\Delta
}_{\mathbf{b;}\overline{\mathbf{g}}}=(1-\Delta _{\mathbf{b;}\overline{\mathbf{g}}});\text{
}\overline{\mathbf{g}}=-\mathbf{g}\ \ .\ \end{array}
\label{eq5}
\end{equation}

In the equations (\ref{eq5}) right side we neglected three-site GF which
give a zero input in a mean field approach with a\ return to GF
$\widehat{Z}_{\mathbf{m}}^{\phi z}$\ , $\widehat{R}_{\mathbf{m;g}}$. It is easy to see
that correlation function
$K_{\widehat{R}_{\mathbf{n}};\widehat{Z}_{\mathbf{r}}^{z\phi }}$\ is a zero if it is calculated using the two- site cluster
state ground state $\Psi _{\mathbf{n,m}}$. After extracting of averages in
GF the right part (\ref{eq5}) the transition to $\mathbf{k}$ representation
gives

\begin{equation}
\begin{array}[t]{c}
(z-2e+\frac{J}{4})\sum_{\mathbf{g}}\langle
\widehat{R}_{\mathbf{k;g}}|\rangle =JP_{\Phi }z_{l}(1-g_{\mathbf{k}})\widetilde{G}_{\mathbf{k}}\ \ \
\end{array}
\label{eq6}
\end{equation}

The equations (\ref{eq3_k})\ and\ (\ref{eq6}) give a final expression for
$\widetilde{G}_{\mathbf{k}}$\

\begin{equation}
\begin{array}{c}
\quad \lbrack
(z-E_{\mathbf{k;}1}^{0})(z-E_{2}^{0})-V_{k}^{2}]\widetilde{G}_{\mathbf{k}}=(z-E_{2}^{0})P_{\phi w};\ \ \ z=\omega +i\delta \\
E_{\mathbf{k;}1}^{0}=e+\frac{J}{4}P_{\phi w}z_{l}g_{\mathbf{k}};\quad
E_{2}^{0}=2e-\frac{J}{4};\quad V_{\mathbf{k}}^{2}=\frac{J^{2}}{16}P_{\Phi
}z_{l}(1-g_{\mathbf{k}}\}.\ \end{array}
\label{eq7}
\end{equation}

Expression $\widetilde{G}_{\mathbf{k}}$\ (\ref{eq7}) describes two triplet
exitations bands $E_{\mathbf{k;\pm
}}=\frac{1}{2}\{(E_{\mathbf{k;}1}^{0}+E_{2}^{0})\pm
\sqrt{(E_{\mathbf{k;}1}^{0}+E_{2}^{0})^{2}+4V_{\mathbf{k}}^{2}}\}$. The lower branch $E_{\mathbf{k;-}}$ (the analogue of $M1$
branch) is close $e$ and has the dispersion with a minimum at AFM vector
$Q$. At $\mathbf{k=Q}$ effective hybridization\ $V_{\mathbf{k}}^{2}$\
(\ref{eq7}) is maximum. Therefore the hybridization between a level $E_{2}^{0}$ and
$E_{\mathbf{k;}1}^{0}$ gives a spectrum of the upper branch $E_{\mathbf{k;+}}$
(analogue of $M3$ branch) with a maximum at $\mathbf{Q}$ (as it is observed
for $M3$ peak dispersion). Our consideration specify that if a system has a
band in a vicinity $e$ then with necessity it has also a band in a vicinity
$2e$.

With temperature increase the intensity of $\chi _{f}(\mathbf{k},\omega )$
decreases due to a factor $P_{\phi w}=\langle \widehat{Z}_{\mathbf{r}}^{\phi
\phi }\rangle -\langle \widehat{Z}_{\mathbf{r}}^{qq}\rangle $: with $T$
-increase the expectation value $\langle \widehat{Z}_{\mathbf{r}}^{\phi \phi
}\rangle $ is decreasing and $\langle \widehat{Z}_{\mathbf{r}}^{qq}\rangle $
increases. In the limit $T\gg e$ the expectation values $\langle
\widehat{Z}_{\mathbf{r}}^{\phi \phi }\rangle $, $\langle
\widehat{Z}_{\mathbf{r}}^{qq}\rangle $ tend to $\frac{1}{4}$\ and the $M1,M3$ -peaks intensity
tends to zero. At the same time it is obviously that with the increase of
the triplet state occupancy $\langle \widehat{Z}_{\mathbf{r}}^{qq}\rangle $
the intensity of quasielastic peak (triplet-triplet transitions) must grow.

Thus, our consideration allows to reflect the basic features of the dynamic
magnetic response in $YbB_{12}$\textbf{\ }\cite{Alekseev07}, first of all
the three-level excitation character with a specific dispersion. In the
conclusion we will notice, that our earlier \ $M1$ peak\ consideration
\cite{Bar09} is close to the present one. But it did not took in the account the
triplet ecitation $\widehat{R}_{\mathbf{n;g}}=i\varepsilon
_{zuq}\widehat{Z}_{\mathbf{n+g}}^{\phi u}\widehat{Z}_{\mathbf{n}}^{\phi q}$, without which it
is impossible to explain the nature of $M3$ peak.

Authors are grateful to P.A.Alekseev and K.S.Nemkovski for useful
discussions.

This work was supported by the Russian Foundation for Basic Research.

\end{document}